\documentclass{article}
\usepackage{amsthm, amssymb, amsmath}
\usepackage{amsthm}
\usepackage{graphicx}
\numberwithin{equation}{section}
\newtheorem{theorem}{Theorem}[section]

\newtheorem{lemma}[theorem]{Lemma}

\newtheorem{proposition}[theorem]{Proposition}

\huge

\newcommand{\lb}{\left(}
\newcommand{\rb}{\right)}

\newcommand{\nc}{\newcommand}

\nc{\be}{\begin{equation}}
\nc{\la}{\label}
\nc{\ba}{\begin{array}}
\nc{\ea}{\end{array}}
\nc{\bs}{\begin{split}}
\nc{\es}{\end{split}}

\nc{\J}{\mathbb J}

\nc{\pt}{\partial_t}
\nc{\ptt}{\partial_t^2}
\nc{\e}{\epsilon}
\nc{\lam}{\lambda}
\nc{\G}{\Gamma}
\nc{\g}{\gamma}
\nc{\al}{\alpha}
\nc{\del}{\delta}
\nc{\Om}{\Omega}

\newcommand{\ra}{\rightarrow}
\renewcommand{\Re}{\operatorname{Re}}
\renewcommand{\Im}{\operatorname{Im}}

\date{}

\newcommand{\DETAILS}[1]{}

\begin{document}
\title{Ballistic Motion of a Tracer Particle Coupled to a Bose gas}
\author{J\"urg Fr\"ohlich\footnote{juerg@phys.ethz.ch; \text{ }present address: School of Mathematics, IAS, Princeton, NJ 08540, USA} \ \text{   }and \ \text{} Zhou Gang\footnote{gangzhou@illinois.edu}}
\maketitle
\setlength{\leftmargin}{.1in}
\setlength{\rightmargin}{.1in}
\normalsize \vskip.1in
\setcounter{page}{1} \setlength{\leftmargin}{.1in}
\setlength{\rightmargin}{.1in}
\large
\centerline{$^{\ast}$Institute of Theoretical Physics, ETH Zurich, CH-8093, Zurich, Switzerland}
\centerline{$^{\dagger}$Department of Mathematics, University of Illinois at Urbana Champaign, U.S.A.}
\date

\setlength{\leftmargin}{.1in}
\setlength{\rightmargin}{.1in} 
\normalsize \vskip.1in
\setcounter{page}{1} \setlength{\leftmargin}{.1in}
\setlength{\rightmargin}{.1in}
\large

\section*{Abstract}
We study the motion of a heavy tracer particle weakly coupled to a dense interacting Bose gas exhibiting Bose-Einstein condensation. In the so-called mean-field limit, the dynamics of this system approaches one determined by nonlinear Hamiltonian evolution equations. We derive the effective dynamics of the tracer particle, which is described by a non-linear integro-differential equation with memory, and prove that if the initial speed of the tracer particle is below the speed of sound in the Bose gas the motion of the particle approaches an inertial motion at constant velocity at large times.
\tableofcontents

\section{Background from Physics and Equations of Motion}
In this paper we study the motion of a very heavy tracer particle coupled to a very dense, very weakly interacting Bose gas at zero temperature, which exhibits Bose-Einstein condensation. In an interacting Bose gas at positive density and zero temperature, the speed of sound is strictly positive. If the initial speed of the tracer particle is well below the speed of sound in the gas one expects that the motion of the particle will approach a uniform ballistic (inertial) motion at large times. A result in this direction has recently been established in a certain limiting regime (the ``mean-field-Bogolubov limit'') of the Bose gas in \cite{EFGSS}. In the present paper, we prove some results complementary to those in \cite{EFGSS} for the same model. (For earlier results on related models, see also
 ~\cite{Spohn2004, KM}.)

We consider a tracer particle of mass $\Lambda M$ and a Bose gas of density $\Lambda \rho_0$ with two-body interactions of strength $\Lambda^{-2} \kappa$. In the so-called mean-field limit of the quantum system, where $\Lambda \rightarrow \infty$, see ~\cite{MR2858064, DFPP}, we obtain the following \textit{classical} equations of motion
\begin{align}
\dot{X_t}=&\frac{P_t}{M},\quad\quad
\dot{P_t}=-\nabla_{X}V(X_t)+g\int dx \text{ }\nabla_{x}W(X_t-x)\lbrace|\alpha_t(x)|^2
-\frac{\rho_0}{g^2}\rbrace,    \label{XPeqns2}\\
i\dot{\alpha}_t(x)=&\lb-\frac{1}{2m}\Delta+gW(X_t-x)\rb\alpha_t(x) 
+\kappa \lb{\phi *\lbrace|\alpha_t|^2-\frac{\rho_0}{g^2}\rbrace}\rb(x)\text{ }\alpha_t(x),              \label{alphaeqn}
\end{align}
In Eqs. \eqref{XPeqns2} and \eqref{alphaeqn}, $X_t\in \mathbb{R}^3$ and $P_{t}\in \mathbb{R}^3$ are the position and momentum of the tracer particle at time $t$, respectively, and $\alpha_t(x)$ is the Ginzburg-Landau order-parameter field describing the state of the Bose gas at time $t$, in the mean-field limit; $W$ and $\phi$ are two-body potentials of short range, with $\phi$ of positive type, and $g$ and $\kappa \geq 0$ are coupling constants. The interpretation of $|\alpha_{t}(x)|^2$ is that of the density of bosonic atoms at the point $x$ of physical space $\mathbb{R}^3$, 
at time $t$. The global phase of $\alpha_t$ is not an observable quantity.
The symbol $*$ in ~\eqref{alphaeqn} denotes convolution.

Eqs. \eqref{XPeqns2} and \eqref{alphaeqn} turn out to be the Hamiltonian equations of motion corresponding to the following Hamilton functional
\begin{align}\label{Ham}
 H(X,P,\alpha,\bar{\alpha})&=&\frac{P^2}{2M}+V(X)+\int dx\text{ } \{
\frac{1}{2m}|\nabla\alpha(x)|^2+gW(X-x)\ (|\alpha(x)|^2-\frac{\rho_0}{g^2})\ \}\\
& &+\frac{\kappa}{2}\int \int dx dy
\ (|\alpha(y)|^2-\frac{\rho_0}{g^2})\ \phi(y-x)
\ (|\alpha(x)|^2-\frac{\rho_0}{g^2}).\nonumber
\end{align}
The Poisson brackets are given by
\begin{equation}\label{PoissonP,X}
\lbrace X^i,X^j\rbrace = \lbrace P_i,P_j \rbrace = 0, \lbrace X^i, P_j \rbrace = \delta^i_j,
\end{equation}
and
\begin{equation}\label{PoissonG-L}
\lbrace \alpha^{\sharp}(x), \alpha^{\sharp}(y)\rbrace = 0, \lbrace \alpha(x), \bar {\alpha}(y) \rbrace = -i \delta (x-y).
\end{equation}

We impose the conditions that $\nabla \alpha_t$ is square-integrable in $x$ and that $|\alpha_t|^2-\frac{\rho_0}{g^2}$ is integrable. In the present paper, we also require that 
$\alpha_{t}(x) \rightarrow \sqrt{\frac{\rho_0}{g^2}}$, as $|x| \rightarrow \infty$. This last condition explicitly breaks 
the gauge invariance, $\alpha^{\sharp}(x) \rightarrow e^{\pm i\theta} \alpha^{\sharp}(x)$, where $\theta$ is an arbitrary angle (independent of $x$ and $t$).
Given our boundary conditions on $\alpha^{\sharp}$ at $\infty$, it is natural to define a new function $\beta$ by
\begin{equation} \label{alphabetarel}
\alpha_t(x)=:\sqrt{\frac{\rho_0}{g^2}}+\beta_t(x),
\end{equation} with $\beta_t(x)\rightarrow 0$, as $|x|\rightarrow \infty.$ 
The equations of motion then read
\begin{align}
\dot{X_t}=&\frac{P_t}{M},\quad\quad
\dot{P_t}=-\nabla_{X}V(X_t)+g\int\nabla_{x}W(X_t-x)\lb|\beta_t(x)|^2
+2\sqrt{\frac{\rho_0}{g^2}}Re\beta_t(x)\rb dx,    \label{XPeqns}\\
i\dot{\beta}_t(x)=&\lb-\frac{1}{2m}\Delta+gW(X_t-x)\rb
\beta_t(x)+\sqrt{\rho_0}W(X_t-x)                  \nonumber\\
+&\kappa\lb\phi *\lb|\beta_t|^2+2\sqrt{\frac{\rho_0}{g^2}}
Re\beta_t\rb\rb(x)\lb\beta_t(x)+\sqrt{\frac{\rho_0}{g^2}}\rb. \label{betaeqn}
\end{align}

The Hamilton functional giving rise to these equations is obtained from \eqref{Ham} after insertion of Eq. 
\eqref{alphabetarel}. It is easy to see that, under rather weak assumptions on the potentials $W$ and $\phi$, Eqs. \eqref{XPeqns} and \eqref{betaeqn} have static solutions, and that if the external force vanishes ($V\equiv 0$) they have ``traveling wave solutions'', provided the speed of the particle is smaller than or equal to the speed of sound in the Bose gas; see \cite{FGSS, EFGSS}. These solutions correspond to an inertial motion of the tracer particle at constant velocity, and the particle is accompanied by a ``splash'' (a coherent cloud) of atoms from the Bose gas. (Quantum mechanically, this splash corresponds to a coherent state of gas atoms 
and causes decoherence in particle-position space, which allows for an essentially ``classical" detection of the particle trajectory.)
If, initially, the speed of the tracer particle is larger than the speed of sound it emits sound waves -- Cherenkov radiation -- into the condensate, which causes friction. As a consequence, the particle loses kinetic energy until its speed has dropped to the speed of sound in the Bose gas. This phenomenon has been analyzed for a simple model (the B-model defined below) in \cite{MR2858064}; (work on Cherenkov radiation in more complicated models is presently carried out).

The following models are of interest (see \cite{FGSS, MR2858064}):
\begin{itemize}
\item[B]-Model: $\kappa=0$ (ideal Bose gas), $g\rightarrow 0,$ see \cite{MR2858064}.
\item[C]-Model: $\kappa=0$, but $g\neq 0;$ see ~\cite{EG2012}.
\item[E]-Model: $2\kappa\rho_0/g^2 :=\lambda=$const., with $g, \kappa\ra  0$ (``Bogolubov limit'').
\item[G]-Model: $\kappa>0$ and $g\neq 0$.
\end{itemize}

In this paper, we focus our attention on the E-model, with $V\equiv 0$ and $\phi(x) = \delta(x)$. The equations of motion then take the form
\begin{align}
\dot{X}_{t}=&\frac{1}{M}P_{t}, \ \dot{P}_{t}=\sqrt{\rho_0}Re\langle \nabla_{x}W^{X_{t}},\ \beta_{t}\rangle \label{eq:momentum}\\
i\dot{\beta}_{t}=&-\frac{1}{2m}\Delta \beta_t+\lambda Re\beta_{t}+\sqrt{\rho_0}W^{X_t}, \label{eq:field}
\end{align}
where
\begin{align}
W^{X}(x):=W(X-x).
\end{align} 
The speed of sound is given by $\sqrt{\frac{\lambda}{2m}}.$

The Hamilton functional of the E-model is
\begin{align}\label{eq:conserv}
H(X, P;\ \bar\beta,\beta):=\frac{|P|^{2}}{2M}+\frac{1}{2m}\int_{\mathbb{R}^3} |\nabla \beta|^2 dx+\lambda \int_{\mathbb{R}^3} |Re\beta|^2 dx+2\sqrt{\rho_0}\int_{\mathbb{R}^3}W^{X}Re \beta dx.
\end{align}

In the subsonic regime the equations of motion have a family of solutions describing inertial particle motion (``traveling-wave solutions'').
The existence and stability of these solutions has been studied in \cite{EFGSS}.
In this paper we extend the results of \cite{EFGSS}. We consider initial conditions at time $t=0$ with the properties that $\beta_0$ is ``small'' in a suitable sense, i.e., the state of the Bose gas is close (or equal) to
the ground state $\beta_0=0$, 
and that the initial speed of tracer particle is well below the speed of sound $\sqrt{\frac{\lambda}{2m}}$. We prove that if the interaction potential $W$ is sufficiently weak the particle will approach an inertial motion 
at a constant speed below the speed of sound, as time $t \rightarrow \infty$.

Our paper is organized as follows: In Section ~\ref{sec:MainTHM}, we describe our main result -- Theorem 2.1 -- which has four parts. Part (1) is proven in Section 3, parts (2) and (3) in Section 4, and the proof is concluded in Section 5. Some technical matters are dealt with in two appendices.

\textit{Notations}: \text{  }By $\mathcal{H}^{k}, \ 1,2,3,\cdots,$ we  denote the Sobolev spaces of complex-valued functions on $\mathbb{R}^{3}$ equipped with the norms
$$\|f\|_{\mathcal{H}^{k}}:=\|(1-\Delta)^{\frac{k}{2}}f\|_2.$$ 
For positive quantities $a$ and $b$, the meaning of ``$a\lesssim b$'' is that there exists a positive constant $C$ such that $a\leq C b$.
The scalar product of two square-integrable functions, $f$ and $g$, on $\mathbb{R}^{3}$ is given by
 $$ \langle f,\ g\rangle:=\int \bar{f}(x) g(x)\ dx.$$
\section*{Acknowledgements}
We are indebted to Daniel Egli, Arick Shao and Israel Michael Sigal for numerous very illuminating
discussions. Our collaboration on the problems solved in this paper has been made possible by a stay at the Institute for Advanced Study in Princeton. We wish to thank our colleagues and the staff at the Institute for hospitality. The stay of J.F. at IAS has been supported by `The Fund for Math' and `The Monell Foundation'.

\section{Statement of the Main Result}\label{sec:MainTHM}
In this section we describe the hypotheses on the potential $W$ and on the choice of initial conditions 
(see hypotheses $(A)$ and $(B)$, below) and state our main results.

\begin{theorem}\label{THM:main}
Suppose the potential $W$ is smooth, spherically symmetric and of rapid decay at spatial $\infty$, and suppose that the positive constants $M$, $m$ and $\lambda$ are of order 1, and the constant $\rho_0>0$ is sufficiently small.

Then there exists a positive constant $\epsilon_0$, such that
\begin{itemize}
\item[(A)] if initially the state of the Bose gas is close to the ground state, $\beta=0$, specifically if $\|(1+|x|^2)^{\frac{5}{2}} \beta_0\|_{\mathcal{H}^3}\leq \epsilon_0$, and
\item[(B)]
if the initial speed of the tracer particle is below the speed of sound in the Bose gas by a sufficient margin, specifically if $|v_0|=\frac{1}{M}|P_0|$ is such that $|v_0|\leq \sqrt{\frac{\lambda}{2m}}-\epsilon_{0},$
\end{itemize}
the following results hold true:
\begin{itemize}
\item[(1)] For any time $t\geq 0$, the speed $|v_{t}|=\frac{|P_{t}|}{M}$ of the tracer particle is strictly below the speed of sound, i.e.,      $|v_{t}|\leq \sqrt{\frac{\lambda}{2m}} - C(\epsilon_0, \rho_0)$, for some $C(\epsilon_0, \rho_0) > 0$.
\item[(2)] At large times, the motion of the tracer particle approaches a uniform ballistic (inertial) motion: There exists some $P_{\infty}\in \mathbb{R}^3$, satisfying $\frac{1}{M}|P_{\infty}|< \sqrt{\frac{\lambda}{2m}}$, such that $P_{t}\rightarrow P_{\infty}$, as $t\rightarrow \infty.$
\item[(3)] In an inertial coordinate system moving with velocity $\frac{1}{M}P_{\infty},$ the particle will come to rest, i.e., there exists $X_{\infty}\in \mathbb{R}^3$ such that $X_{t}-\frac{1}{M}P_{\infty} t\rightarrow X_{\infty}$, as $t\rightarrow \infty,$
\item[(4)] The wave function $\beta_t$ describing the state of the Bose gas approaches a ``traveling wave'' accompanying the particle, in the sense that there exists a function $\beta_{\infty}\in L^{2}(\mathbb{R}^3)$ with the property that
\begin{align*}
\|(1+|x-X_t|)^{-1}[\beta_t-\beta_{\infty}(\cdot-X_t)]\|_{\infty}\rightarrow 0,\ \text{as}\ t\rightarrow \infty.
\end{align*}
\end{itemize}
\end{theorem}
This theorem is proven in the remainder of our paper. 

As a preliminary to the proof of Theorem \ref{THM:main}, one must establish local and global wellposedness of the equations of motion \eqref{eq:momentum} and \eqref{eq:field}. This is easy, because, for an arbitrary given particle trajectory 
$\lbrace X_t \rbrace_{0 \leq t < \infty}$, the equation for $\beta_t$, i.e., Eq. \eqref{eq:field}, is linear.  
Details can be found in \cite{EFGSS} and will not be repeated here. 

\section{An Application of Energy Conservation: Proof of Part (1)}
To prove part (1) of Theorem 2.1, we make use of energy conservation, which holds because \eqref{eq:momentum} and \eqref{eq:field} are the equations of motion of an autonomous Hamiltonian system.

The only non-positive term in the Hamilton functional \eqref{eq:conserv} is \text{ }$2\sqrt{\rho_0} \int_{\mathbb{R}^3} W^X Re\beta\ dx$. 
This term can be bounded by applying the Schwarz inequality, which yields
\begin{align}
2\sqrt{\rho_0} |\int_{\mathbb{R}^3} Re\beta W^{X} dx| &\leq  \lambda \int_{\mathbb{R}^3} (Re\beta)^2 dx+\frac{4\rho_0}{\lambda} \int_{\mathbb{R}^3} (W^{X})^2 dx \nonumber\\
&=\lambda \int_{\mathbb{R}^3} (Re\beta)^2 dx+\frac{4\rho_0}{\lambda} \int_{\mathbb{R}^3} W^2 dx.\label{eq:holder}
\end{align}
Putting this estimate back into the Hamilton functional $H$ (see \eqref{eq:conserv}), and using our assumptions on the initial conditions, we obtain
\begin{align*}
H(X_0, P_0,\beta_0,\bar\beta_{0} )\leq &\frac{|P_0|^2}{2M}+\frac{1}{2m}\int_{\mathbb{R}^3} |\nabla \beta_0|^2 dx+2\lambda \int_{\mathbb{R}^3} |Re\beta_0|^2 dx+\frac{4\rho_0}{\lambda} \int_{\mathbb{R}^3} W^2 dx.
\end{align*}
For any time $t$, Eqs. \eqref{eq:conserv} and
\eqref{eq:holder} also yield the lower bound
$$H(X_t, P_t,\beta_t,\bar\beta_{t} )\geq \frac{|P_{t}|^2}{2M}-\frac{4\rho_0}{\lambda} \int_{\mathbb{R}^3} W^2 dx.$$
These upper and lower bounds, together with \textit{energy conservation}, i.e., $H(X_t, P_t,\beta_t,\bar\beta_{t} )=H(X_0, P_0,\beta_0,\bar\beta_{0} )$, imply that there is a constant $c>0$ such that, for any time $t$,
\begin{align*}
|P_{t}|^2\leq |P_0|^2+c(\|\beta_0\|_{H^1}^2+\rho_0).
\end{align*}

Hence if $v_{\max}:=\frac{1}{M}[|P_0|^2+c(\|\beta_0\|_{H^1}^2+\rho_0)]^{\frac{1}{2}}< \sqrt{\frac{\lambda}{2m}}$ then it follows that, for any time $t$,
\begin{align}
|v_{t}|=\frac{|P_t|}{M}\leq v_{\max}< \sqrt{\frac{\lambda}{2m}}\text{ },\label{eq:subsonicM}
\end{align}
 i.e., the speed of the tracer particle remains strictly in the subsonic regime, for all times $t$.\\
 This proves Part (1) of Theorem ~\ref{THM:main}.

\section{Effective Dynamics of the Tracer Particle: Proofs of Parts (2) and (3)}
To start with our proofs of parts (2) and (3) of Theorem \ref{THM:main} we recast the equations of motion in a convenient form. We observe that equation \eqref{eq:field} is merely real-linear, rather than complex-linear, in $\beta_t$, and hence it is convenient to rewrite it as a system of equations for $\text{Re}\beta_t$ and $\text{Im}\beta_t$.

We define a vector function ${\bf{h}}_{t}:\ \mathbb{R}^3\rightarrow \mathbb{R}^2$ by
$${\bf{h}}_{t}  (x-X_{t})=\left[
\begin{array}{lll}
\text{Re}\beta_{t}(x)\\
\text{Im}\beta_{t}(x)
\end{array}
\right].$$
Then ~\eqref{eq:momentum} and ~\eqref{eq:field} become
\begin{align}
\dot{X}_{t}=&\frac{1}{M}P_{t}, \ \dot{P}_{t}=\sqrt{\rho_0}\langle
\left[
\begin{array}{ccc}
\nabla_{x}W\\
0
\end{array}
\right],
 \bf{h}_{t}\rangle,\label{eq:traj1}\\
\dot{\bf{h}}_{t}=&H(t) {\bf{h}}_{t}-\sqrt{\rho_0} \left[
\begin{array}{lll}
0\\
W
\end{array}
\right], \label{eq:field1}
\end{align}
where $H(t)$ is the $2\times 2$ matrix operator given by
\begin{align}
H(t):=\left[
\begin{array}{ccc}
\frac{1}{M}P_{t}\cdot \nabla_{x}& -\frac{1}{2m}\Delta\\
-(-\frac{1}{2m}\Delta+\lambda) & \frac{1}{M}P_{t}\cdot\nabla_{x}
\end{array}
\right].
\end{align}

To further simplify the equations, we define a new vector function, $\bf{\delta}_{t}$, by
\begin{align}
{\bf{h}}_{t}=\sqrt{\rho_0}H^{-1}(t) \left[
\begin{array}{lll}
0\\
W
\end{array}
\right]+\bf{\delta}_t.\label{eq:decom1}
\end{align} 
Rewriting Eqs. \eqref{eq:traj1} and \eqref{eq:field1} in terms of $\bf{\delta}_t$, we find the equations
\begin{align}
\dot{X}_{t}=&\frac{1}{M}P_{t}, \ \dot{P}_{t}=\sqrt{\rho_0}\langle
\left[
\begin{array}{ccc}
\nabla_{x}W\\
0
\end{array}
\right],
 \bf{\delta}_{t}\rangle\label{eq:traj2},\\
\dot{\bf{\delta}}_{t}=&H(t) {\bf{\delta}}_{t}+\frac{\sqrt{\rho_0}}{M} H^{-2}(t) \dot{P_t}\cdot \nabla_{x}\left[
\begin{array}{lll}
0\\
W
\end{array}
\right].\label{eq:field2}
\end{align}
The initial condition for $\delta$ is given by
\begin{align}\label{eq:delta0}
\delta_0=-\sqrt{\rho_0} H^{-1}(0)\left[
\begin{array}{lll}
0\\
W
\end{array}
\right]+\bf{h}_0,
\end{align} and, in the equation for $\dot{P}_t$, we have used the fact that 
$\left[
\begin{array}{ccc}
\nabla_{x}W\\
0
\end{array}
\right] \perp H^{-1}(t) \left[
\begin{array}{lll}
0\\
W
\end{array}
\right].$ (To verify this claim, we determine the explicit expression for $H^{-1}(t)$ and use the fact that $W$ is spherically symmetric -- the details are straightforward.)

Applying Duhamel's principle to equation \eqref{eq:field2} we find that
\begin{align}
{\bf{\delta}_{t}}=&U(t,0){\bf{\delta}_0}  +\frac{\sqrt{\rho_0}}{M} \int_{0}^{t} U(t,s)  H^{-2}(s) \dot{P_s}\cdot \nabla_{x}\ ds  \left[
\begin{array}{lll}
0\\
W
\end{array}
\right]\nonumber\\
=&U(t,0) {\bf{h}_0} -\sqrt{\rho_0}U(t,0) H^{-1}(0)\left[
\begin{array}{lll}
0\\
W
\end{array}
\right]+\frac{\sqrt{\rho_0}}{M} \int_{0}^{t} U(t,s)  H^{-2}(s) \dot{P_s}\cdot \nabla_{x}\ ds  \left[
\begin{array}{lll}
0\\
W
\end{array}
\right],\label{eq:expreDelta}
\end{align}
where $U(t,s)$ is the propagator (from time $s$ to time $t$) generated by the time-dependent operator $H(\cdot)$,  and we have used the expression for $\delta_0$ in \eqref{eq:delta0}.

We now insert Eq. \eqref{eq:expreDelta} into the equation of motion \eqref{eq:traj2} for $P_t$ to obtain
\begin{align}
\dot{P}_t
=&\sqrt{\rho_0}\langle
\left[
\begin{array}{lll}
\nabla_{x}W\\
0
\end{array}
\right],
 U(t,0) {\bf{h}_0}\rangle\nonumber\\
&-\rho_0 \langle
\left[
\begin{array}{lll}
\nabla_{x}W\\
0
\end{array}
\right],
U(t,0) H^{-1}(0)\left[
\begin{array}{ccc}
0\\
W
\end{array}
 \right]\rangle
 \nonumber\\
&+\frac{\rho_0}{M}\langle
\left[
\begin{array}{lll}
\nabla_{x}W\\
0
\end{array}
\right],\ \int_{0}^{t} U(t,s) H^{-2}(s) \dot{P_s}\cdot \nabla_{x}\ ds  \left[
\begin{array}{lll}
0\\
W
\end{array}
\right]
\rangle\nonumber\\
=:& D_1+D_2+D_3,\label{eq:D3}
\end{align}
where the terms $D_{k},\ k=1,2,3,$ correspond to the first, second and third line, respectively, on the right side. These terms depend on the velocity $\frac{1}{M}P_t$, which, by  \eqref{eq:subsonicM}, remains subsonic. In order to have a quantitative measure by how much $\frac{1}{M}|P_{t}|$ remains below the speed of sound 
$\sqrt{\frac{\lambda}{2m}}$, we introduce the quantity
\begin{align}\label{eq:subsonicL}
\eta:=\sup_{t\in [0,\infty)}\frac{\frac{|P_t|}{M}}{\sqrt{\frac{\lambda}{2m}}}.
\end{align}
The results in Sect. 3 show that $0\leq \eta<1$.

The key result of the present section is the following lemma.
\begin{lemma}\label{LM:est} There exists a constant $C(\eta)$, which is uniformly bounded on any closed subset of $[0,1]$ not containing 1, such that,
\begin{align}
|D_1(t)|\leq C(\eta) (1+t)^{-3}\sqrt{\rho_0} \|(1+x^2)^{2} {\bf{h}}_0\|_{\mathcal{H}^2},\label{eq:estD1}
\end{align}
\begin{align}
|D_2(t)|\leq C(\eta) \rho_0 (1+t)^{-3},
\end{align} and
\begin{align}
|D_3(t)|\leq C(\eta) \rho_0 G(t) (1+t)^{-3}.
\end{align}
\end{lemma}
Here the function $G$ is defined as
\begin{align}\label{eq:majorant}
G(t):=\max_{0\leq s\leq t}(1+s)^3 |\dot{P}_s|.
\end{align}

The proof of this lemma is contained in Appendix ~\ref{sec:A}.

We now return to equation \eqref{eq:D3} for $\dot{P}_t$. By definition, the function $G(t)$ is increasing in $t$, which implies that, for any $t\geq 0$,
\begin{align}
G(t)\leq C(\eta) [\sqrt{\rho_0} \|(1+x^2)^{3} {\bf{h}}_0\|_{H^2}+\rho_0+\rho_0 G(t)].
\end{align}
Thus, if $\rho_0$ is so small that $C(\eta) \rho_0\leq \frac{1}{2}$ then
\begin{align*}
G(t)\leq 2C(\eta) [\sqrt{\rho_0} \|(1+x^2)^{3} {\bf{h}}_0\|_{\mathcal{H}^2}+\rho_0]
\end{align*} which, in view of the definition of $G$, implies that 
\begin{align}
|\dot{P}_t|\leq C(1+t)^{-3},\label{eq:ptdecay}
\end{align}
for any time $t\geq 0$, where $C$ is the finite constant given by the right side of the inequality above.

Next, we show how Parts (2) and (3) of Theorem 2.1 follow from \eqref{eq:ptdecay}.

Since, by \eqref{eq:ptdecay}, $\dot{P}_t$ is integrable on $[0,\infty)$, we conclude that there exists a $P_{\infty}\in \mathbb{R}^3$ such that $P_t\rightarrow P_{\infty}$, as $t\rightarrow \infty.$ This is Part (2) of Theorem 2.1.

The position, $Y_t$, of the tracer particle in the inertial coordinate system moving with velocity $\frac{1}{M}P_{\infty}$ is given by
$$Y_t:=X_t-t\frac{P_{\infty}}{M}.$$ 
Its derivative, the velocity of the tracer particle in the moving coordinate system, is obtained from
\begin{align*}
\dot{Y}_t=\frac{1}{M}[P_t-P_{\infty}]=-\frac{1}{M}\int_{t}^{\infty}\dot{P}_s\ ds.
\end{align*} 
Together with the estimate in \eqref{eq:ptdecay}, this identity implies that
$$|\dot{Y}_t|\lesssim C(1+t)^{-2},$$
i.e., $\dot{Y}_t$ is integrable in $t$, which implies that there exists some $Y_{\infty}$ such that $Y_{t}\rightarrow Y_{\infty}$, as $t\rightarrow \infty$. This proves Part (3) of Theorem 2.1.

\section{The State of the Bose Gas, as $t\rightarrow \infty$: Proof of Part (4)}

In our proof of Part (3) of Theorem 2.1, the function 
${\bf{h}}_{t}:=\left[
\begin{array}{lll}
\text{Re}\beta_t (x+X_t)\\
\text{Im}\beta_t (x+X_t)
\end{array}
\right]$ 
has been written as a sum of two terms, 
$\sqrt{\rho_0} H^{-1}(t) \left[
\begin{array}{lll}
0\\
W
\end{array}
\right]$ 
and $\delta_t$; see Eq. \eqref{eq:decom1}. Hence, in order to establish Part (4) of Theorem 2.1, it suffices to prove that 
\begin{align}
\|(1+|x|)^{-1} \delta_t\|_{\infty}\rightarrow 0,\label{eq:fir}
\end{align}
as $t\rightarrow \infty$, with 
$\beta_{\infty}= \sqrt {\rho_0} H^{-1}(\infty) 
\left[
\begin{array}{lll}
0\\
W
\end{array}
\right].$

As follows from Eq. \eqref{eq:expreDelta} for $\delta_t$, we only need to estimate the decay in time of the following three vector functions: $U_1(t):=U(t,0){\bf{h}_0}$, \text{   } $U_2(t):=U(t,0) H^{-1}(0)\left[
\begin{array}{ccc}
0\\
W
\end{array}
\right]$,
\text{   } and $U_3(t,s):=U(t,s) H^{-2}(s)\left[
\begin{array}{ccc}
0\\
\nabla W
\end{array}
\right].$
  
In these estimates it is used that the potential $W$ is smooth and of rapid decay at spatial infinity.

\begin{proposition}\label{Prop:estimates}
\begin{align}
\|(1+|x|)^{-1}U_1(t)\|_{\infty}\lesssim (1+t)^{-1} \|(1+x^2)^{3}{\bf{h}_0}\|_{H^2}.
\end{align}
\begin{align}
\|(1+|x|)^{-1}U_2(t)\|_{\infty}\lesssim (1+t)^{-1},
\end{align}
and
\begin{align}
\|(1+|x|)^{-1}U_3(t,s)\|_{\infty}\lesssim (1+t-s)^{-1}.
\end{align}
\end{proposition}
This proposition is proven in Appendix B.

Next, we use Proposition \ref{Prop:estimates} to complete our proof of decay estimates on $\delta_t$ in a weighted $L^{\infty}$-space, thus establishing Part (4) of Theorem 2.1. 
Taking the $\|(1+|x|)^{-1}(\cdot)\|_{\infty}$ - norm of both sides of Eq. \eqref{eq:expreDelta} and then using Proposition \ref{Prop:estimates}, we conclude that
\begin{align*}
\|(1+|x|)^{-1}\delta_{t}\|_{\infty}\lesssim &\sum_{k=1}^{2} \|(1+|x|)^{-1} U_k(t)\|_{\infty}+\int_{0}^{t} \|(1+|x|)^{-1} U_3(t,s)\|_{\infty} |\dot{P}_s|\ ds\\
\lesssim & (1+t)^{-1}+\int_{0}^{t}(1+t-s)^{-1} |\dot{P}_s|\ ds\\
\lesssim & (1+t)^{-1}
\end{align*} 
which is the desired estimate.

\appendix

\section{Proof of Lemma ~\ref{LM:est}}\label{sec:A}
In what follows we start by recasting the expressions appearing in Lemma 4.1 in a convenient form. Subsequently we sketch the main ideas of the proof.

A key element of the proof will be to understand the propagator $U(t,s)$ from time $s$ to time $t$ generated by 
$H(t)$. The operator $H(t)$ is the sum of two terms
$$H(t)=H_0+\frac{P_{t}}{M}\cdot \nabla_{x},$$ 
where $H_{0}$ is given by
\begin{align*}
H_0:=\left[
\begin{array}{ccc}
0& -\frac{1}{2m}\Delta\\
-(-\frac{1}{2m}\Delta+\lambda) &0
\end{array}
\right].
\end{align*} 
The observation that these two terms commute enables us to derive an explicit expression for $U(t, s)$:
\begin{align}
U(t,s) =e^{H_0(t-s)} e^{(X_t-X_s)\cdot\nabla_{x}}.
\end{align}
In our calculations, it is useful to diagonalize the matrix operator $H_0$ and hence the operator $e^{H_0(t-s)}$, which, by  Eq. (A.1), leads to the diagonalisation of the propagator $U(t,s).$

To simplify our notations, we use units such that
\begin{align}
2m=\lambda=1.
\end{align}

We define a $2\times 2$ matrix operator, $A$, by
\begin{align}\label{eq:difA}
A:=\left[
\begin{array}{ccc}
\sqrt{-\Delta} & \sqrt{-\Delta}\\
i\sqrt{-\Delta+1} & -i\sqrt{-\Delta+1}
\end{array}
\right].
\end{align} 
By Fourier transformation and direct computation we find that
$$A^{-1}=-\frac{1}{2i  }L^{-1}\left[
\begin{array}{ccc}
-i\sqrt{-\Delta+1} & -\sqrt{-\Delta}\\
-i\sqrt{-\Delta+1} & \sqrt{-\Delta}
\end{array}
\right]$$
and
\begin{align*}
A^{-1} e^{H_0(t-s)} A
=&\left[
\begin{array}{ccc}
e^{i L (t-s)} &0\\
0& e^{-i L (t-s)}
\end{array}
\right]
\end{align*} 
with $L$ given by
$$L:=\sqrt{-\Delta+1}\sqrt{-\Delta}.$$

Inserting one of the identities $A^{-1}A=AA^{-1}= Id_{2\times 2}$ at appropriate places in the terms $D_1$, $D_2$ and $D_3$ of Eq. \eqref{eq:D3}, we obtain the expressions
\begin{align}
D_1(t)=-\text{Re}\langle\sqrt{-\Delta}  \nabla_{x}W, \ i e^{i t L } e^{(X_t-X_0)\cdot \partial_{x}} L^{-1} [i\sqrt{-\Delta+1} \ \text{Re}\beta_0+\sqrt{-\Delta}\ \text{Im}\beta_0]\rangle,
\end{align}
\begin{align}
D_2(t)=-\rho_0 \text{Re}\langle \sqrt{-\Delta}  \nabla_{x}W,\ e^{i t L } e^{(X_t-X_0)\cdot \nabla_{x}} [L-i \frac{P_0}{M}\cdot \partial_{x}]^{-1}[-\Delta+1]^{-\frac{1}{2}} W
\rangle,
\end{align} 
and
\begin{align}
D_3(t)=-\frac{\rho_0}{M} \text{Re}\ i\langle \sqrt{-\Delta}  \nabla_{x}W,\ \int_{0}^{t} e^{i (t-s) L } e^{(X_t-X_s)\cdot \nabla_{x}}[L-i \frac{P_s}{M}\cdot \partial_{x}]^{-2}[-\Delta+1]^{-\frac{1}{2}} P_s\cdot \nabla_{x}W\ ds\rangle.
\end{align}

To prove decay estimates on these terms we use that $e^{i (t-s) L } e^{(X_t-X_s)\cdot \nabla_{x}}$ and $e^{i t L } e^{(X_t-X_0)\cdot \nabla_{x}}$ are oscillatory operators in momentum space. More precisely,
we Fourier-transform the functions inside the scalar products on the right sides of Eqs. (A.4) - (A.6), and then employ standard integration-by-parts techniques.

We start with analyzing $D_1$. Without loss of generality, we assume that
\begin{align}
(X_t-X_0)\cdot \nabla_{x}= |X_t-X_0|\partial_{x_3}.
\end{align} 
After Fourier transformation, and by introducing polar coordinates, we obtain
\begin{align*}
D_1(t)=&Re \int_{0}^{2\pi} \int_{0}^{\pi}\int_{0}^{\infty} e^{it \rho [\sqrt{1+\rho^2}-\frac{|X_t-X_0|}{t} cos\theta]}\rho^3 F_1(\rho,\theta,\alpha)\ d\rho d\theta d\alpha\\
&+ Re \int_{0}^{2\pi} \int_{0}^{\pi}\int_{0}^{\infty} e^{it \rho [\sqrt{1+\rho^2}-\frac{|X_t-X_0|}{t} cos\theta]}\rho^4 F_2(\rho,\theta,\alpha)\ d\rho d\theta d\alpha,
\end{align*} 
where $F_1:= sin\theta\ \overline{\Hat{W}}\ \Hat{\Re\beta_0}\ n(\theta,\alpha)$ and $F_2:=-i \frac{sin\theta\ \overline{\hat{W}}\ \hat{\Im \beta_0}}{\sqrt{1+\rho^2}}\ n(\theta,\alpha)$, and $n$ is the unit vector
\begin{align}\label{eq:defF}
n(\theta,\alpha):=(cos\alpha sin\theta, sin\alpha sin\theta, cos\theta)^{T}
\end{align} 
Obviously $F_1$ and $F_2$ are complex-valued functions whose smoothness and decay rate at 
$\rho=\infty$ are determined by those of $W$, $\text{Re}\beta_0$ and $\text{Im}\beta_0$.

To prove an appropriate decay estimate we integrate by parts in the variable $\rho$,
using the recipe
\begin{align}
e^{it \rho [\sqrt{1+\rho^2}-\frac{|X_t-X_0|}{t} cos\theta]}=\frac{1}{it}\frac{1}{ \sqrt{1+\rho^2}+\frac{\rho^2}{\sqrt{1+\rho^2}}-\frac{|X_t-X_0|}{t} cos\theta }\partial_{\rho}e^{it \rho [\sqrt{1+\rho^2}-\frac{|X_t-X_0|}{t} cos\theta]}.\label{eq:recipe}
\end{align}
The denominator, $\sqrt{1+\rho^2}+\frac{\rho^2}{\sqrt{1+\rho^2}}-\frac{|X_t-X_0|}{t} cos\text{ }\theta$, on the right side can be seen to be strictly and uniformly positive
by using the observation that, for $\eta<1$ (see ~\eqref{eq:subsonicL}), we have that
\begin{align}
\frac{|X_t-X_0|}{t}cos\text{ }\theta \leq \eta < 1,
\end{align} 
as follows by inserting \eqref{eq:subsonicL} into $\frac{|X_t-X_0|}{t}=\frac{1}{t}|\int_{0}^{t}\frac{P_s}{M}\ ds|$. (We note that the speed of sound, $\sqrt{\frac{\lambda}{2m}}$, is $=1$, because we have assumed that 
$2m=\lambda=1$.) Consequently
\begin{align}\label{eq:lower}
0<\frac{1}{ \sqrt{1+\rho^2}+\frac{\rho^2}{\sqrt{1+\rho^2}}-\frac{|X_t-X_0|}{t} cos\theta }\leq \frac{1}{1-\eta}.
\end{align}

The factors $\rho^3$, $\rho^4$, respectively, appearing in the integrands of the two terms for $D_1(t)$ enable us to integrate by parts three times, the boundary terms being zero, and we find that
$$|D_1(t)|\lesssim t^{-3}C(\eta)\|(1+|x|^2)^2 \beta_0\|_{\mathcal{H}^2},$$ 
where $C(\eta)$ comes from controlling the denominator in \eqref{eq:recipe} (see also ~\eqref{eq:lower}). It is quite tedious, albeit not difficult, to understand how the factor
 $\|(1+|x|^2)^2 \beta_0\|_{\mathcal{H}^2}$ emerges from these calculations. We skip details and refer to a previous paper \cite{EFGSS}, where a very similar problem has been dealt with in a general setting.

A direct argument shows that
$$|D_1(t)|\lesssim C(\eta)\|(1+|x|^2)^2 \beta_0\|_{\mathcal{H}^2},$$
which implies that our estimates do not blow up at $t=0$.

The last two estimates clearly imply the desired bound \eqref{eq:estD1}.

\textit{Bound on} $D_2(t)$: After Fourier transformation and introduction of polar coordinates, as in our analysis of $D_1(t)$, we obtain that
\begin{align*}
D_2(t)=\rho_0 \text{Re}\ i \int_{0}^{2\pi}\int_{0}^{\pi}\int_{0}^{\infty} e^{it \rho [\sqrt{1+\rho^2}-\frac{|X_t-X_0|}{t} cos\theta]} \rho^3 F(\rho^2,\theta,\alpha)\ d\rho d\theta d\alpha.
\end{align*} 
The key observation is that the function $$F:=[\sqrt{1+\rho^2}-\frac{P_0}{M} \cdot f(\theta,\alpha)]^{-1} (1+\rho^2)^{-1}|\hat{W}|^2 n(\theta,\alpha),$$ with $n(\theta,\alpha)$ defined in ~\eqref{eq:defF}, is a \textit{real-valued, smooth} function of \textit{fast decay}, which follows from the fact that, under our hypotheses on the potential $W$, $|\hat{W}|^2$ is real, spherically symmetric, smooth (in $\rho^2$) and of fast decay. Thanks to the presence of ``$\text{Re}\ i$'' in front of the integrations on the right side, we may extend the range of $\rho$ from $[0,\infty)$ to $(-\infty,\infty)$ and obtain
\begin{align}\label{eq:end}
D_2(t)=\frac{1}{2}\rho_0 \text{Re}\ i \int_{0}^{2\pi}\int_{0}^{\pi}\int_{-\infty}^{\infty} e^{it \rho [\sqrt{1+\rho^2}-\frac{|X_t-X_0|}{t} cos\theta]} \rho^3 F(\rho^2,\theta,\alpha)\ d\rho d\theta d\alpha.
\end{align} This enables us to use the recipe in \eqref{eq:recipe} to integrate by parts as many times as needed, which then yields the desired estimate
\begin{align}
|D_2(t)|\lesssim \rho_0 C(\eta) (1+t)^{-3}.
\end{align}

Finally, we turn to the proof of our bound on  the term $D_3$, which has the form
\begin{align}\label{eq:D33}
D_3(t)=\rho_0 \int_0^{t}K(t,s)\cdot P_s\ ds.
\end{align} 
By arguments very similar to those used in the analysis of $D_2$, before \eqref{eq:end}, we find that
$$K(t,s)=Re\  \int_{0}^{2\pi}\int_{0}^{\pi}\int_{-\infty}^{\infty} e^{i(t-s) \rho [\sqrt{1+\rho^2}-\frac{|X_t-X_s|}{t-s} cos\theta]} \rho^2 F_2(\rho^2,\theta,\alpha)\ d\rho d\theta d\alpha,$$ 
where $F_2$ is a real-valued, smooth function of rapid decay, found after a similar analysis as in $D_2$. Here we rotate the axis to make $(X_t-X_s)\cdot \partial_{x}$ to be $|X_t-X_s|\partial_{x_3}$.

As in our analysis of $D_2$, we integrate by parts three times and then apply a direct estimate to obtain 
\begin{align}
|K(t,s)|\lesssim C(\eta) (1+t-s)^{-3}.
\end{align}
Plugging this into ~\eqref{eq:D33} we find that
\begin{align*}
|D_3(t)|\lesssim &\rho_0 C(\eta)\int_0^{t}(1+t-s)^{-3} |\dot{P}_s|\ ds\\
\lesssim &\rho_0 C(\eta) \int_{0}^{t}(1+t-s)^{-3} (1+s)^{-3}\ ds\ G(t)\\
\lesssim &\rho_0 C(\eta)(1+t)^{-3} G(t),
\end{align*} 
with $G$ as in \eqref{eq:majorant}.

\section{Proof of Proposition ~\ref{Prop:estimates}}\label{sec:B}
The idea underlying our proof is not very difficult. We exploit the dispersive nature of the propagators $U(t,s)$ to prove that $\delta_t$ tends to $0$, as $t \rightarrow \infty$, in a weighted $L^{\infty}-$ norm.
In what follows, we discuss this idea in some detail.

As in Appendix ~\ref{sec:A}, we assume that $2m=\lambda=1$. 

We start by deriving simple expressions for $U(t,s)H^{-n}(s), \ n=0,1,2.$
The matrix $A$ defined in ~\eqref{eq:difA} can be used to diagonalize these three operators:
\begin{align}
&U(t,s)H^{-n}(s)\\
=&A A^{-1} U(t,s) A [A^{-1} H(s) A]^{-n} A^{-1}\nonumber\\
=&A \left[
\begin{array}{ccc}
e^{i(t-s)L} e^{(X_t-X_s)\cdot \nabla_{x}} [iL+\frac{P_s}{M}\cdot\partial_{x}]^{-n} & 0\\
0 & e^{-i(t-s)L} e^{(X_t-X_s)\cdot \partial_{x}} [-iL+\frac{P_s}{M}\cdot\nabla_{x}]^{-n}
\end{array}
\right] A^{-1}\nonumber\\
=&\left[
\begin{array}{ccc}
K_1(t,s,n), & K_2(t,s,n)\\
K_3(t,s,n), & K_1(t,s,n)
\end{array}
\right]
\end{align}
with $$K_1:=
\frac{1}{2}\{e^{i(t-s)L} e^{(X_t-X_s)\cdot \nabla_{x}} [iL+\frac{P_s}{M}\cdot\nabla_{x}]^{-n}+e^{-i(t-s)L} e^{(X_t-X_s)\cdot \nabla_{x}} [-iL+\frac{P_s}{M}\cdot\nabla_{x}]^{-n}\},$$
\begin{align*}
K_2:=&\frac{\sqrt{-\Delta}}{2i \sqrt{-\Delta+1}}e^{i(t-s)L} e^{(X_t-X_s)\cdot \nabla_{x}} [iL+\frac{P_s}{M}\cdot\nabla_{x}]^{-n}\\
&-\frac{\sqrt{-\Delta}}{2i \sqrt{-\Delta+1}}e^{-i(t-s)L} e^{(X_t-X_s)\cdot \nabla_{x}} [-iL+\frac{P_s}{M}\cdot\nabla_{x}]^{-n}
\end{align*}
and
\begin{align*}
K_3:=&-\frac{\sqrt{-\Delta+1}}{2i \sqrt{-\Delta}}e^{i(t-s)L} e^{(X_t-X_s)\cdot \nabla_{x}} [iL+\frac{P_s}{M}\cdot\nabla_{x}]^{-n}\\
&+\frac{\sqrt{-\Delta+1}}{2i \sqrt{-\Delta}}e^{-i(t-s)L} e^{(X_t-X_s)\cdot \nabla_{x}} [-iL+\frac{P_s}{M}\cdot\nabla_{x}]^{-n}.
\end{align*}

Among all the terms to be analyzed we choose to study the following one
\begin{align}
D:=\frac{\sqrt{-\Delta+1}}{\sqrt{-\Delta}} e^{itL} e^{(X_t-X_0)\cdot \nabla_{x}}Re\beta_{0},
\end{align} 
which is a component of the vector function $U_1(t)$ introduced in Sect. 5. All other terms can be estimated in a very similar way, the estimates being somewhat easier. 

We propose to prove the estimate
\begin{align}\label{eq:desired}
\|(1+|x|)^{-1}D\|_{\infty} \lesssim \|(-\Delta+1)^{\frac{3}{2}} (1+|x|^2)^{\frac{5}{2}}Re\beta_0\|_{2}.
\end{align}

By Fourier transformation we have that
$$D=\frac{1}{(2\pi)^3} \int_{\mathbb{R}^3} e^{ik\cdot x} \frac{\sqrt{2+|k|^2}}{|k|} e^{it[|k|\sqrt{1+|k|^2}-\frac{|X_t-X_0|}{t}k_3]}\ \widehat{Re\beta_0}\ d^3 k,$$
Introducing polar coordinates we obtain
$$D=C\int_{0}^{2\pi}\int_{0}^{\pi}\int_{0}^{\infty} e^{i\rho f(\theta,\alpha)\cdot x} e^{it\rho[ 1+ \sqrt{1+\rho^2}-\frac{|X_t-X_0|}{t} cos\theta]}\rho  F(\rho,\theta,\alpha)\ d\rho d\theta d\alpha ,$$
for some constant $C\in \mathbb{C}, $where $F= \sqrt{1+\rho^2}\ sin\text{ }\theta\ \widehat{Re\beta_{0}}$, and $f(\theta,\alpha)\in \mathbb{R}^3$ is a real vector.

To derive the desired decay estimate we integrate by parts, using ~\eqref{eq:recipe}, and then take absolute values. We find that
\begin{align*}
|D|\lesssim &\frac{ (1+|x|)}{t}\int_{0}^{2\pi}\int_{0}^{\pi}\int_{0}^{\infty} \frac{1}{1+\rho}[|\partial_{\rho} F|(\rho,\theta,\alpha)+| F|(\rho,\theta,\alpha)]\ \ d\rho d\theta d\alpha\\
\lesssim & \frac{ (1+|x|)}{t} [\|(1+\rho^2)^{\frac{3}{2}}|\partial_{\rho} F|(\rho,\theta,\alpha)\|_{\infty}+\|(1+\rho^2)^{\frac{3}{2}}| F|(\rho,\theta,\alpha)\|_{\infty}]\\
\lesssim & \frac{ (1+|x|)}{t} [\sum_{k=1}^{3} \|(-\Delta+1)^{\frac{3}{2}} x_{k}Re\beta_0\|_{1}+\|(-\Delta+1)^{\frac{3}{2}}Re\beta_0\|_{1}]\\
\lesssim & \frac{ (1+|x|)}{t}\|(-\Delta+1)^{\frac{3}{2}} (1+|x|^2)^{\frac{5}{2}}Re\beta_0\|_{2}.
\end{align*} 
By a direct estimate on $D$, it is seen that our estimates do not blow up at $t=0$: We have that
\begin{align*}
|D|\lesssim \|(-\Delta+1)^{\frac{3}{2}} (1+|x|^2)^{\frac{5}{2}}Re\beta_0\|_{2}.
\end{align*} 
Together with the estimate above, this implies \eqref{eq:desired}, which is the desired result.
\def\cprime{$'$} \def\cprime{$'$} \def\cprime{$'$} \def\cprime{$'$}
  \def\cprime{$'$} \def\cprime{$'$}

\end{document}